# Strong suppression of near-field thermal transport between twisted bilayer graphene near the magic angle

Fuwei Yang[1,2] and Bai Song[1,3,4*]

[1]*Beijing Innovation Center for Engineering Science and Advanced Technology, Peking University, Beijing 100871, China.*

[2]*Center for Nano and Micro Mechanics, Tsinghua University, Beijing 100084, China.*

[3]*Department of Energy and Resources Engineering, Peking University, Beijing 100871, China.*

[4]*Department of Advanced Manufacturing and Robotics, Peking University, Beijing 100871, China.*

*Corresponding author. Email: songbai@pku.edu.cn

**ABSTRACT:** Twisted bilayer graphene (TBLG) has recently emerged as a versatile platform for studying a variety of exotic transport phenomena. Here, we present a theoretical study of near-field thermal radiation between suspended TBLG with a focus on the magic angle. Within the chirally symmetric continuum model, we observe a suppressed heat flow when approaching the magic angle owing to a reduced Drude weight, with greater suppressions at lower temperatures and larger gap sizes. When the chemical potential lies in the energy gap near the charge neutrality point, more than 100-fold heat-flow variation can be achieved at 50 K within 0.25° of twist. By reducing the electron scattering rate, the radiation spectrum near the magic angle dramatically narrows, leading to over 10,000-fold of suppression. In addition, supported TBLG is briefly considered to facilitate experimental measurement. With rationally tailored substrates, the heat-flow contrast can still exceed 1000. We also discuss lattice relaxation effect in terms of the interlayer coupling energy, finding that a stronger coupling leads to a smaller heat-flow contrast and more prominent multiband transport. Our results highlight the great potential of magic-angle TBLG in thermal transport, especially for controlling thermal radiation.

# I. INTRODUCTION

The interlayer twist angle in bilayer graphene offers an unprecedented degree of freedom for tuning a broad variety of material properties and physical phenomena. To name a few, twist-dependent electronic structures [1–3], optical transitions [4,5], mechanical friction [6,7], thermal conductivity [8–11], and chemical reactivity [12,13] have all been predicted and demonstrated in twisted bilayer graphene (TBLG). In particular, the so-called magic twist angle of about 1° has recently sparked considerable interest. Inspired by the unexpected observation of superconducting and correlated insulating states [14,15], a range of other exotic phenomena have subsequently been discovered in magic-angle twisted bilayer graphene (MATBLG), such as charge-ordered states [16,17], quantized anomalous Hall effect [18–20], and the Pomeranchuk effect [21,22]. Underlying these phenomena is the unique electronic structure of MATBLG, which features a pair of remarkably flat bands and a vanishing Fermi velocity near the Dirac points [3]. In light of the close relation between electronic structures and optical emissions, thermal radiation of MATBLG is also expected to be of great interest but has remained largely unexplored, especially in the near field.

Near-field radiative heat transfer (NFRHT) can exceed Stefan-Boltzmann's blackbody limit to far-field radiation by orders of magnitude [23–31], and offers rich opportunities for actively controlling and harnessing thermal radiation in various energy conversion and heat management devices [32–37]. Previous work has shown that for relatively large angles ($\geq 1.5°$), a small interlayer twist around some special angles can tune the heat flow between two TBLG sheets by over an order of magnitude at room temperature [38]. In the proximity of the magic angle, lattice relaxation strongly modifies the band structure of TBLG and leads to the opening of a low-energy gap near the $\Gamma$ point [39–41]. In addition, the surface plasmon polaritons (SPPs) in MATBLG shows a rather flat dispersion, which is distinct from that of large-angle TBLG [42–45]. Therefore, NFRHT between MATBLG may be considerably more interesting and deserves a systematic study.

Here, we theoretically study near-field thermal radiation between two suspended TBLG sheets with a focus on the magic angle (Fig. 1). We describe the electronic properties of TBLG using a continuum model, and capture the lattice relaxation effect by varying the interlayer coupling energy, which intuitively reflect variations in the interlayer distance across the AA and AB/BA stacking regions [Fig. 1(b)]. Small chemical potentials and low temperatures are of particular interest since the flat bands and the band gap near the magic angle both appear at low energies. Employing fluctuational electrodynamics, we first show that the radiative heat flow is quickly suppressed as the magic angle is approached at room temperature. This suppression increases with decreasing temperature, and reaches over two orders of magnitude at 50 K for micrometer gaps due to a vanishing Drude weight. More importantly, by reducing the electron scattering rate, over 10,000-fold heat flow suppression can be achieved. From an experimental point of view, we also propose substrates that can potentially maintain a large contrast. We conclude with a discussion of the lattice relaxation effect both on the total heat flow and on the radiation spectrum.

## II. METHODS

A detailed description of the methods can be found in our previous work [38]. Here we only provide a brief summary and focus on the special treatment near the magic angle. Within the framework of fluctuational electrodynamics, the total heat flux between two parallel planes across a vacuum gap *d* is given by [26]

$$q(T_1, T_2, d) = \int_0^\infty d\omega [\Theta(\omega, T_1) - \Theta(\omega, T_2)] \, f(\omega). \tag{1}$$

Here, $f(\omega) = \int_0^\infty dk \frac{k}{4\pi^2} [\tau_s(\omega, k) + \tau_p(\omega, k)]$ is the spectral transfer function, and $\Theta$ is the average energy of a harmonic oscillator. *T*, *ω*, and *k* are the temperature, frequency, and parallel wavevector component, respectively. $\tau_s$ and $\tau_p$ represent the photon transmission probabilities for the *s*- and *p*-polarized waves. In the linear regime, we define the spectral and total heat transfer coefficient as $h_\omega = \frac{\partial \Theta(\omega, T)}{\partial T} f(\omega)$ and $h = \int_0^\infty d\omega h_\omega$, respectively. For TBLG, $\tau_p$ dominates NFRHT and is often calculated from the optical conductivity [38].

Using the Kubo formula, we first compute the real part of the optical conductivity $\sigma$ of TBLG from its electronic band structure [46,47]. The imaginary part is then obtained with the Kramers-Kronig relation. For graphene, $\sigma$ is the sum of an intraband (Drude) term $\sigma_D$ and an interband term $\sigma_I$, with $\sigma_D$ given by

$$\sigma_D(\omega) = \frac{D}{\pi} \frac{i}{\hbar\omega + i\Gamma}. \tag{2}$$

Here, $\Gamma$ is the electron scattering rate and is set as 7 meV unless otherwise noted [46]. *D* is the Drude weight characterizing the strength of intraband transitions [48].

The low-energy electronic structure of TBLG is well described by an effective continuum model [3,40,47,49], which captures the long-range moiré modulation of the interlayer coupling strength and employs the following Hamiltonian:

$$H = \begin{pmatrix} H_1 & U \\ U^\dagger & H_2 \end{pmatrix}. \tag{3}$$

Here, $H_l$ ($l$ = 1, 2) is the Dirac Hamiltonian for the two layers given by

$$H_l(\mathbf{k}) = \hbar v_F (\mathbf{k} - \mathbf{K}_l) \cdot \boldsymbol{\sigma}, \tag{4}$$

where $\boldsymbol{\sigma}$ denotes the Pauli matrices and $v_F = \sqrt{3} a_0 t_0 / 2\hbar$ is the Fermi velocity, with $a_0$ being the lattice constant and $t_0$ the intralayer coupling energy set as 2.7 eV [43,50,51]. Assuming nearest-neighbor coupling [$\mathbf{q}_j$ in Fig. 1(c)], the interlayer hopping term $U$ can be written as [3,40]

$$U = u' \sum_{j=1}^{3} \exp(-i\mathbf{q}_j \cdot \mathbf{r}) U^j, \tag{5}$$

with $U^1 = \begin{pmatrix} u/u' & 1 \\ 1 & u/u' \end{pmatrix}$, $U^2 = \begin{pmatrix} ue^{i\phi}/u' & 1 \\ e^{-i} & ue^{i\phi}/u' \end{pmatrix}$, and $U^3 = \begin{pmatrix} ue^{-i\phi}/u' & 1 \\ e^{i\phi} & ue^{-i\phi}/u' \end{pmatrix}$.

Here, $u$ and $u'$ denote the coupling energies in the AA and AB/BA regions, respectively, and $\phi = \frac{2}{3}\pi$.

The effect of lattice relaxation is considered important for twist angles below 2°, and is usually incorporated by setting $u \neq u'$ [39,40]. Theoretically, $u$ = 79.7 meV and $u'$ = 97.5 meV are predicted [40] and commonly used [43,52,53]. Experimental results are relatively scarce, with a recent measurement obtaining a smaller $u$ of ~40 meV [45]. The discrepancy is attributed to electron-electron interactions and extrinsic factors in sample preparation [45], suggesting that $u$ is potentially tunable. In particular, the chirally symmetric model with $u$ = 0 is of great interest, as it concisely captures the fundamental features of MATBLG including the vanishing Fermi velocity and flattening of the lowest bands [54]. Note that since the band structure only depends on $u/u'$, below we fix $u'$ at 97.5 meV and consider the effect of varying $u$.

## III. RESULTS AND DISCUSSION

### A. Band structure near the magic angle

The largest twist angle leading to either a vanishing Fermi velocity or a bandwidth minimum is usually defined as the first magic angle, or simply the magic angle [54]. Here, we adopt the latter definition which can be expressed as [54]

$$\theta_{\text{magic}} = 2\arcsin\left(\frac{3u'}{4\sqrt{3}\pi t_0 \times 0.586}\right). \tag{6}$$

With $t_0$ = 2.7 eV and $u'$ = 97.5 meV, the magic angle is $\theta_{\text{magic}}$ = 0.97°. In Fig. 2(a), we plot the band structures of relaxed MATBLG for both the chirally symmetric model ($u$ = 0) and $u$ = 79.7 meV, together with the case when lattice relaxation is neglected ($u$ = 97.5 meV). The characteristic flat bands are clearly seen for all three models near the K (K') point [Fig. 2(a)], which notably remains flat through the entire Brillouin zone for $u$ = 0 [54]. In addition, an energy gap opens at the Γ point as a result of lattice relaxation. Increasing $u$ decreases the gap and shifts the corresponding interband transitions to lower energies. As the twist angle increases to 3° and further to 5°, the influence of lattice relaxation moves to higher energies and the low-energy bands remain similar, as shown in Fig. S1 of the supplemental material [55].

Focusing on the chirally symmetric model, we further plot the density of states (DOS) at and near $\theta_{\text{magic}}$ in Fig. 2(b), including $\theta$ = 1.05° and 1.35°. Around the charge neutrality point (CNP), the perfectly flat bands at the magic angle leads to a sharp van-Hove singularity in the DOS, which splits into two symmetric peaks as the twist angle slightly increases. Immediately adjacent to these peaks is a region of negligible DOS, which corresponds to the low-energy gap discussed earlier. These characteristics near the CNP are expected to strongly affect the SPPs of TBLG and consequently NFRHT.

### B. Thermal transport at room temperature

We first consider radiative thermal transport at room temperature and zero chemical potential. The latter condition ($\mu = 0$) facilitates manifestation of the intriguing electronic structures of near-magic-angle TBLG around the CNP. In Fig. 3, we plot the total heat transfer coefficient (HTC) and the corresponding Drude weight ($D$) as a function of the twist angle. The results are generally similar to the case of $\mu = 0.25$ eV discussed in previous work without considering lattice relaxation [38]. Briefly, at large angles, the HTC approaches that of the decoupled bilayer graphene for both a 10 nm gap and a 1 μm gap. As the twist angle decreases, the HTC at the 1 μm gap decreases together with $D$, while the HTC at the 10 nm gap increases [38]. Despite these similarities, the 10 nm-gap HTC clearly differs from previous work as the twist angle approaches $\theta_{magic}$, which decreases with $D$ and reaches a minimum instead of a maximum around the special angle where $D$ is a minimum [38].

To understand such a difference, we plot in Fig. 4(a) the photon transmission probability at $\theta_{magic}$ to analyze the characteristics of the $p$-polarized SPPs. In contrast to the case of $\mu = 0.25$ eV where the intraband SPPs span the entire energy range of interest to NFRHT at room temperature [38], a large gap of negligible transmission dominates the spectrum for $\mu = 0$, while the intraband plasmons appear rather flat and are confined to a very narrow region below the gap. Similar features are expected also for TBLG near the magic angle, albeit at higher energies (Fig. S2 [55]). Therefore, as $D$ decreases with reducing $\theta$, the low-transmission gap—instead of the SPPs that enhance HTC for $\mu = 0.25$ eV—shifts to lower energy that is more important to thermal transport, leading to the suppression of HTC. In addition, we note that the strong interband plasmons above the gap also add to the heat flow, with a larger contribution at smaller $\theta$, which will be discussed below in more details. The combined contribution of the intraband and interband plasmons thus leads to the observation of a HTC minimum in Fig. 3.

### C. Enhanced heat-flow contrast at low temperatures

To further explore the potential of MATBLG, we analyze thermal radiation at a lower temperature with a thermal energy that better matches the few-meV width of the flat bands and the width of the adjacent gap on the order of 10 meV (Fig. 2). For a quick comparison with the room-temperature scenario, we plot in Fig. 4(b) the photon transmission probability $\tau_p$ at $T = 100$ K and $\mu = 0$. The interband plasmons show little difference while the intraband plasmons experience a stronger confinement at lower $T$, together with a widened gap. These features in the SPPs suggest a more dramatic heat-flow suppression at lower temperature as $\theta$ approaches the magic angle. In Fig. 5(a), we show the HTC at 100 K, which varies by roughly 12- and 43-fold within 0.25° of twist for a 10 nm and a 1 μm gap, respectively, while the HTC at 300 K only varies by about 4- and 15-fold for over 1° of twist (Fig. 3). As expected, the enhanced contrasts at lower temperature are induced by a steeper drop of the Drude weight [Fig.

5(b)]. In Fig. 5(c), we plot the ratio of HTC for $\theta = 10°$ ($h_{10°}$) and $\theta = 0.97°$ ($h_{0.97°}$) for large gaps and observe that the contrast remains greater than 10 for gaps up to ~10 μm. Small gaps are of less interest as shown in Fig. S3 [55], and will not be the focus below.

In order to further enhance the heat-flow contrast, we first increase the maximum heat flow by increasing the chemical potential from 0 to 0.05 eV. This leads to an increase of the maximum $D$ and a shift of the special angle for minimum $D$ to 1.05° [Fig. 5(b)]. Correspondingly, the heat-flow contrast $h_{10°} / h_{1.05°}$ increases regardless of the gap size and reaches 2 orders of magnitude for gaps from 1 μm to 6 μm [Fig. 5(c)]. Nevertheless, increasing $\mu$ also increases the minimum $D$ [Fig. 5(b)], which tends to lower the contrast. To suppress the minimum heat flow, we drop the temperature to 50 K which reduces the minimum $D$ by over 100 times without affecting the maximum $D$ [Fig. 5(b)]. As a result, the heat-flow contrast remains greater than 100 for $d > 2$ μm [Fig. 5(c)]. We note that the separate control of the maximum and minimum $D$ is possible for two reasons. On one hand, the chemical potential $\mu = 0.05$ eV lies inside the energy gap for TBLG near the magic angle, which should lead to zero Drude weight if no thermal broadening is considered. On the other hand, the Drude weight of large-angle TBLG shows a negligible temperature dependence for $\mu \gg k_B T$, similar to the case of monolayer graphene [55].

The above discussion shows the possibility of drastically suppressing the heat flow via a vanishing Drude weight for TBLG near the magic angle at specific chemical potentials and temperatures. In Fig. 5(d), we plot the $T$-dependence of $D$ at three representative $\mu$ for the special angle $\theta = 1.05°$, in order to further reveal the interplay among the temperature, chemical potential, and the characteristic energy gap. For $\mu = $ 0 and 0.002 eV, the Drude weight differs only at temperatures below ~30 K, because both chemical potentials lie within the narrow band around the CNP. In both cases, $D$ varies non-monotonically with $T$, and thus reducing $T$ may not lead to a larger variation in the heat flow. When $\mu$ lies instead in the gap ($\mu = 0.05$ eV), $D$ increases monotonically with $T$, showing a typical thermal activation behavior. Therefore, as long as $T$ is sufficiently low, $D$ would be too small to contribute to thermal transport, and decreasing the temperature further would not lead to stronger suppression.

### D. Potential for 10000-fold heat-flow suppression

In addition to the temperature and chemical potential, the carrier-scattering rate $\Gamma$ in graphene can vary by orders of magnitude and provides yet another parameter for tuning NFRHT [34,56,57]. For monolayer graphene, reducing $\Gamma$ leads to a blue-shift of the spectral transfer function and a narrower peak, which is characteristic of intraband plasmons [56,57]. These features are also expected for large-angle TBLG sheets since they can be approximated by two decoupled monolayers. In Fig. 6(a), we show an example with $\theta = 10°$ by reducing $\Gamma$ from 7 meV to 0.7 meV. Intriguingly, for $\theta = 1.05°$, the opposite trend is observed, with a smaller $\Gamma$ resulting in a red-shifted spectrum [Fig. 6(b)]. Moreover, the peak width is considerably reduced by a factor of 8.9, as compared to 1.3 times for $\theta = 10°$. These differences indicate a substantially different transport mechanism near the magic angle. Indeed, due to the extremely small Drude weight, the imaginary part of the optical conductivity is negative. Consequently, intraband plasmons can no longer be supported and contribute to thermal radiation [38,46,48].

The dramatic variation of the spectral transfer function near the magic angle combined with a moderate change at large angles leads to a stronger suppression of the heat flow with smaller scattering rate, as demonstrated in Fig. 6(c) at $T = 50$ K and $\mu = 0.05$ eV. Notably, the heat-flow contrast exceeds 10,000 at gap sizes around 10 μm, with $h_{10°}$ remaining above or comparable to the blackbody limit. To experimentally observe such strong suppression, substrate-supported TBLG can be helpful. As discussed in previous work [38], the substrate generally reduces the heat-flow variation with the twist angle. Fortunately, it is still possible to maintain a rather high contrast by carefully tailoring the substrate. Here, we put the TBLG sheet on a thin dielectric film with a vacuum-like permittivity and a thick layer of reflective metal on the backside. The former suppresses the broadband propagating modes, while the latter isolates the system from environmental radiation [58]. With a 1-μm-thick film of silicon carbide (SiC) and bulk silver (Ag), $h_{10°} / h_{1.05°}$ can still reach 100 when $\Gamma = 7$ meV. If $\Gamma$ is reduced to 0.7 meV, over 1000-fold contrast can be achieved [Fig. 6(d)]. Note the choice of the substrate materials and thicknesses of each layer is rather flexible [58–60].

### E. Effect of varying the interlayer coupling energy

Focusing on the chirally symmetric model ($u = 0$) for MATBLG, we have now demonstrated the possibility to tune NFRHT by four orders of magnitude at tailored chemical potentials, sufficiently low temperatures, and relatively large gaps. In this section, we systematically study the effect of lattice relaxation by comparing three representative interlayer coupling energies of 0, 79.7 meV, and 97.5 meV. The condition of room temperature and zero chemical potential is considered to highlight some of the key characteristics.

Varying $u$ causes substantial variations in the low-energy electronic structures of MATBLG [Fig. 2(a)] but trivial changes for large twist angles (Fig. S1). In particular, the energy gap near the CNP narrows with increasing $u$ and the low-energy DOS increases. Given its close correspondence with the DOS [46,47], the Drude weight shows distinct behaviors near the magic angle—with larger $u$ leading to larger $D$, while approaching a constant value regardless of $u$ at large angles [Fig. 7(a)]. Correspondingly, the HTCs at large angles are indistinguishable for different $u$ [Fig. 7(b)-(c)]. At small angles, the HTCs for a 1 μm gap follows the trend of the Drude weight with a reduced heat-flow contrast as $u$ increases, while those for a 10 nm gap show slight deviations due to non-negligible interband contributions.

As shown in previous work, the interband transitions shift to low energies and become increasingly more important to NFRHT at small gap sizes as the twist angle decreases to ~2°. In addition, multiple peaks in radiation spectrum induced by interband plasmons become more visible especially at elevated temperatures [38]. Similar features are also expected for TBLG near the magic angle. To demonstrate the effect of $u$ in this aspect, we plot in Fig. 7(d)-(f) the spectral heat transfer coefficient (sHTC) at a small gap of 10 nm. Despite the difference in the details of the spectra, some general features are consistent for different $u$. First, the radiation spectra show a strong multiband feature for $u$ = 79.7 and 97.5 meV even at room temperature, which is relatively weak for the case of $u$ = 0. Additionally, the peak heights at relatively high energies increase as the twist angle decreases for all three cases. Both observations are due to a red-shift of the interband transitions either with increasing $u$ or decreasing $\theta$.

With a large $u$, the radiation spectra consist mainly of peaks from interband plasmons as $\theta$ approaches $\theta_{\text{magic}}$, which leads to the non-monotonic variation of the HTC at the 10 nm gap [Fig. 7(b)-(c)] and its slight deviation from the trend of the Drude weight.

## IV. CONCLUSION

In summary, we theoretically demonstrate the strong suppression of near-field radiative thermal transport between two suspended twisted bilayer graphene sheets near the magic angle. With a sufficiently low temperature, a tailored chemical potential, and a small carrier-scattering rate, over 10,000-fold heat-flow suppression can be achieved as the twist angle is reduced to the vicinity of the magic angle. This giant contrast is predicted for the conditions of large vacuum gaps and temperatures considerably higher than the liquid helium temperature, both of which are helpful for experimental measurement. To this end, we have also offered a set of commonly used substrates that can maintain a large contrast of over 1000. To study the lattice relaxation effect, we consider three representative interlayer coupling energies, and find that a stronger coupling results in a smaller heat-flow contrast and more prominent multiband features in the radiation spectrum. Our results provide useful insights into thermal transport of magic-angle twisted bilayer graphene, and highlight its potential for manipulating radiative heat flow. Nevertheless, our current work is still based on a single-electron picture and limited to relatively high temperatures. The implications of strong correlation in MATBLG on radiative thermal transport may represent an interesting topic for future exploration.

# ACKNOWLEDGEMENT

This work was supported by the National Natural Science Foundation of China (Grant No. 52076002), the Beijing Innovation Center for Engineering Science and Advanced Technology, the XPLORER PRIZE from the Tencent Foundation, the Tsien Excellence in Engineering program, and the High-performance Computing Platform of Peking University. We thank Qizhang Li, Lanyi Xie, and Haiyu He for helpful discussions.

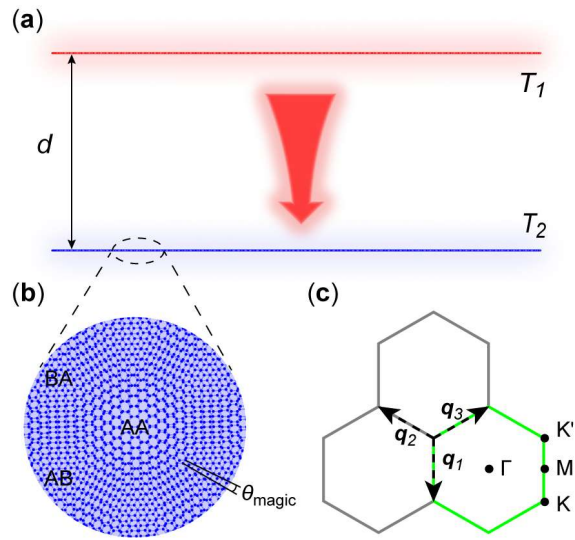

FIG. 1. Schematic of radiative thermal transport between two identical TBLG sheets separated by a vacuum gap. (a) Side view of the system. (b) Top view of a TBLG sheet at the magic angle with the AA and AB/BA stacking regions marked. (c) The reciprocal space with the green hexagon showing the moiré Brillouin zone and $q_j$ representing the nearest-neighbor interlayer hopping.

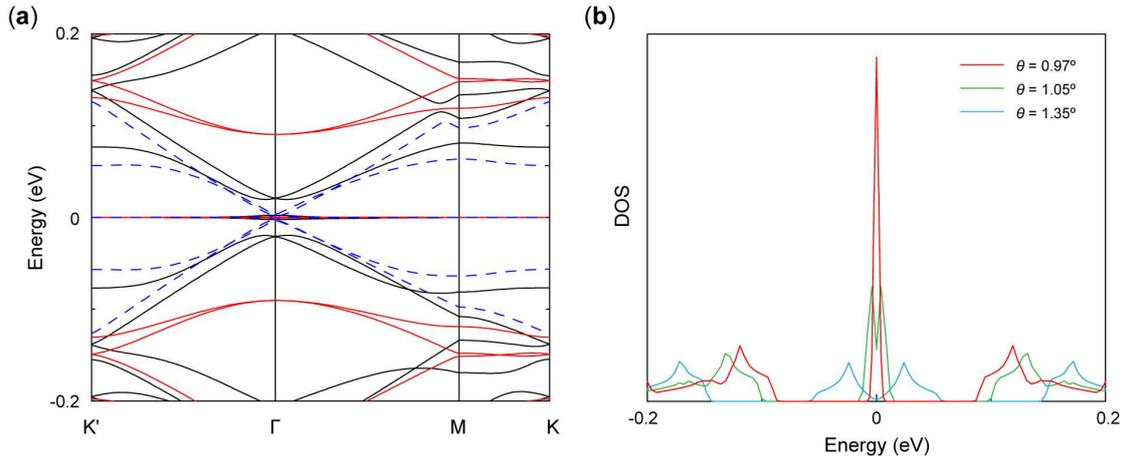

FIG. 2. Electronic structures of TBLG near the magic angle. (a) Calculated band structure at the magic angle $\theta = 0.97°$ with $u = 0$ (red solid line), 79.7 meV (black solid line), and 97.5 meV (blue dashed line, only the six lowest bands are shown). (b) The density of states of TBLG with $\theta = 0.97°$, 1.05°, and 1.35° for $u = 0$.

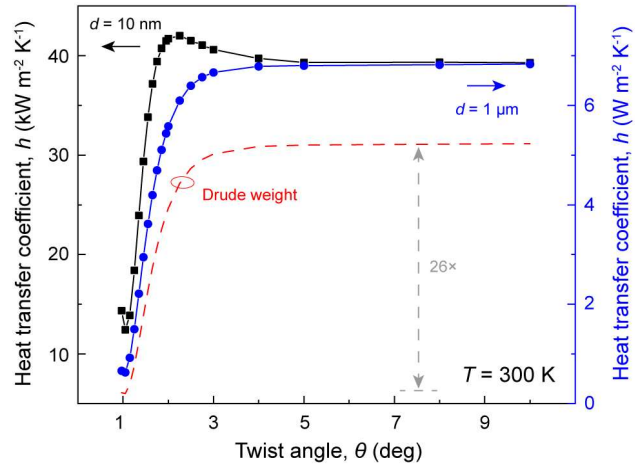

FIG. 3. Heat transfer coefficient as a function of the twist angle at $T = 300$ K and $\mu = 0$ for $d = 10$ nm and 1 μm, together with the Drude weight in arbitrary unit.

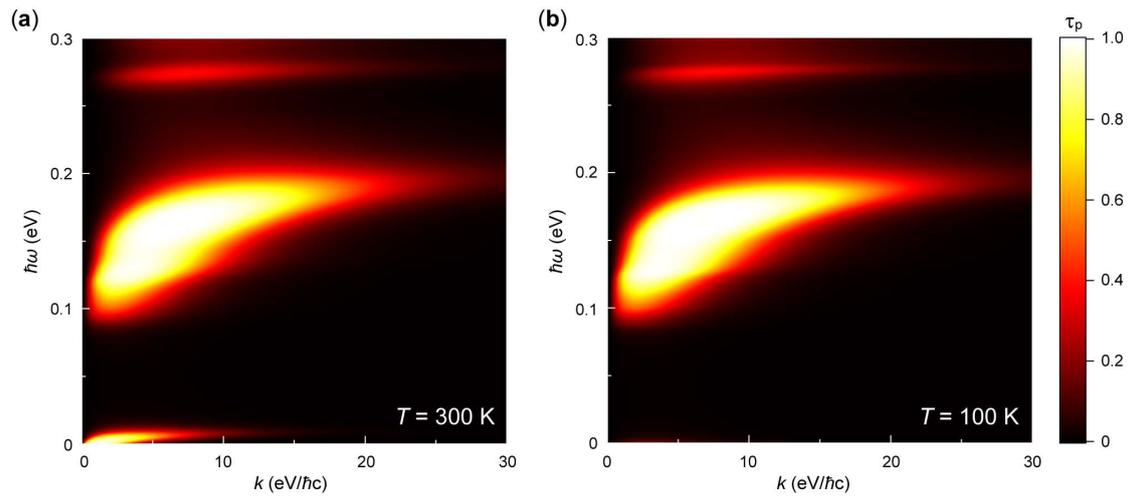

FIG. 4. Photon transmission probability $\tau_p$ of TBLG with $\theta = 0.97°$ and $\mu = 0$ across a 10 nm gap at (a) $T = 300$ K and (b) 100 K.

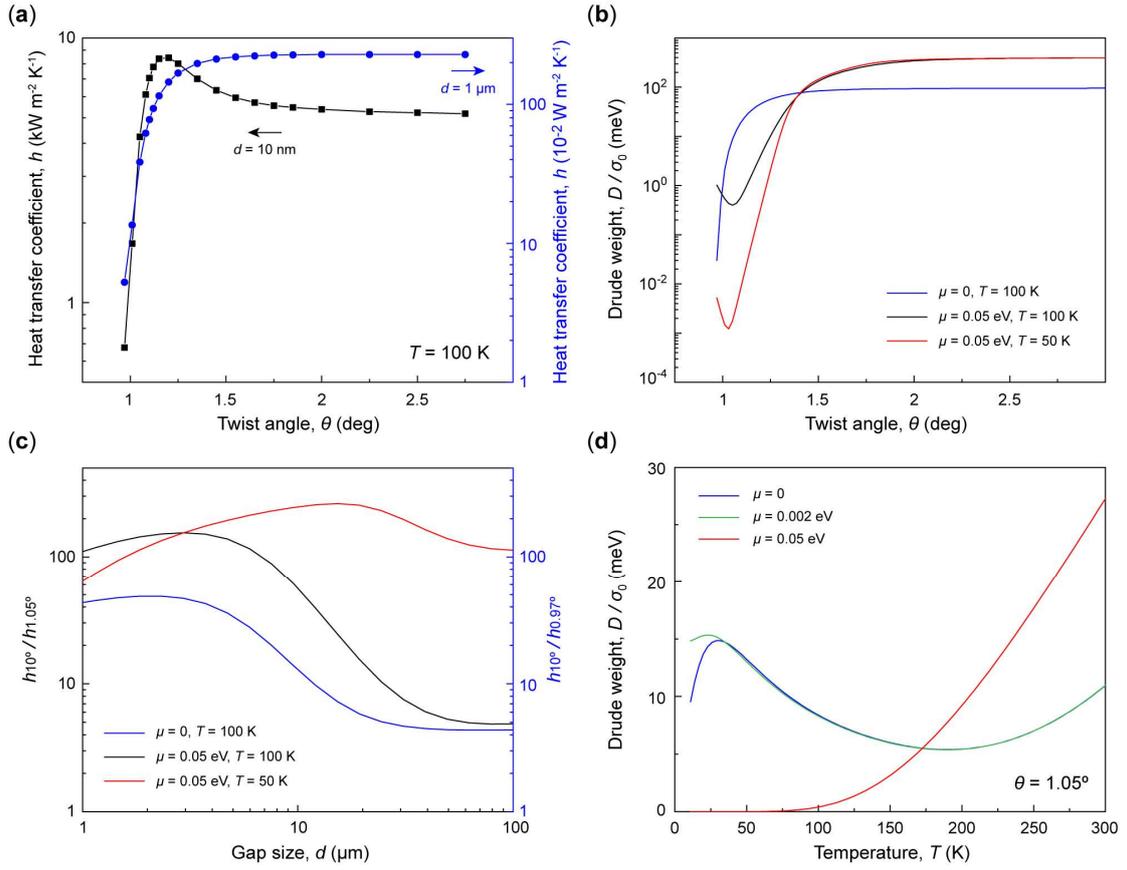

FIG. 5. Enhanced heat-flow contrast at low temperature. (a) Heat transfer coefficient as a function of twist angle at $T = 100$ K and $\mu = 0$ for $d = 10$ nm and 1 μm. (b) The Drude weight at select temperatures and chemical potentials. $\sigma_0 = e^2/4\hbar$ is the universal conductivity of monolayer graphene. (c) Increased heat-flow contrast at $T = 50$ K and $\mu = 0.05$ eV. (d) The variation of the Drude weight with temperature for $\theta = 1.05°$ at representative chemical potentials.

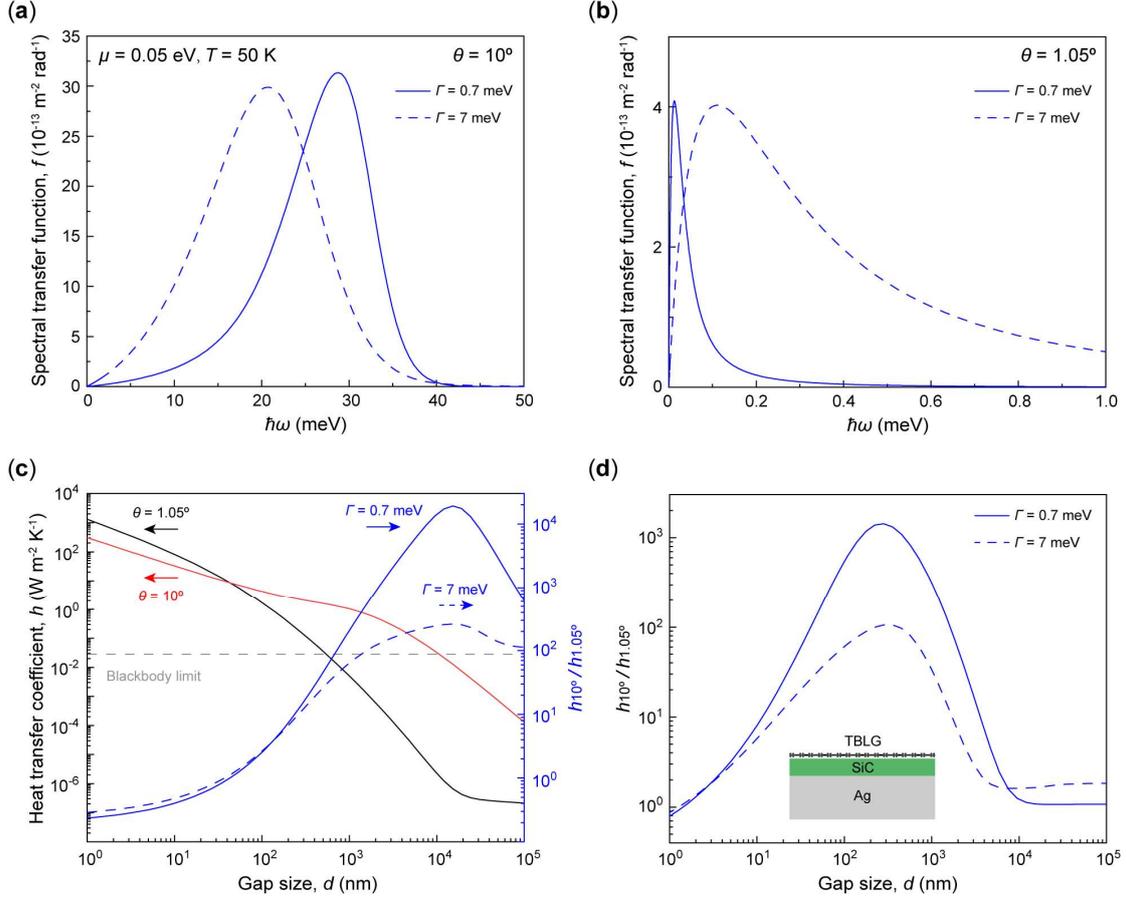

FIG. 6. Potential for substantial heat-flow suppression near the magic angle at cryogenic temperatures. The spectral transfer function comparing $\Gamma = 7$ meV and 0.7 meV at $T = 50$ K, $\mu = 0.05$ eV, and $d = 1$ μm, for TBLG with (a) $\theta = 10°$ and (b) $\theta = 1.05°$. (c) The effect on the corresponding heat flow. The left axis shows the heat transfer coefficient as a function of gap size with $\Gamma = 0.7$ meV, and the right axis shows the ratio of $h_{10°}$ to $h_{1.05°}$. (d) The effect of tailored substrate. The inset shows a schematic of one side, with TBLG on a 1 μm-thick SiC film supported by a bulk Ag layer. The permittivities are given in the supplemental material [55].

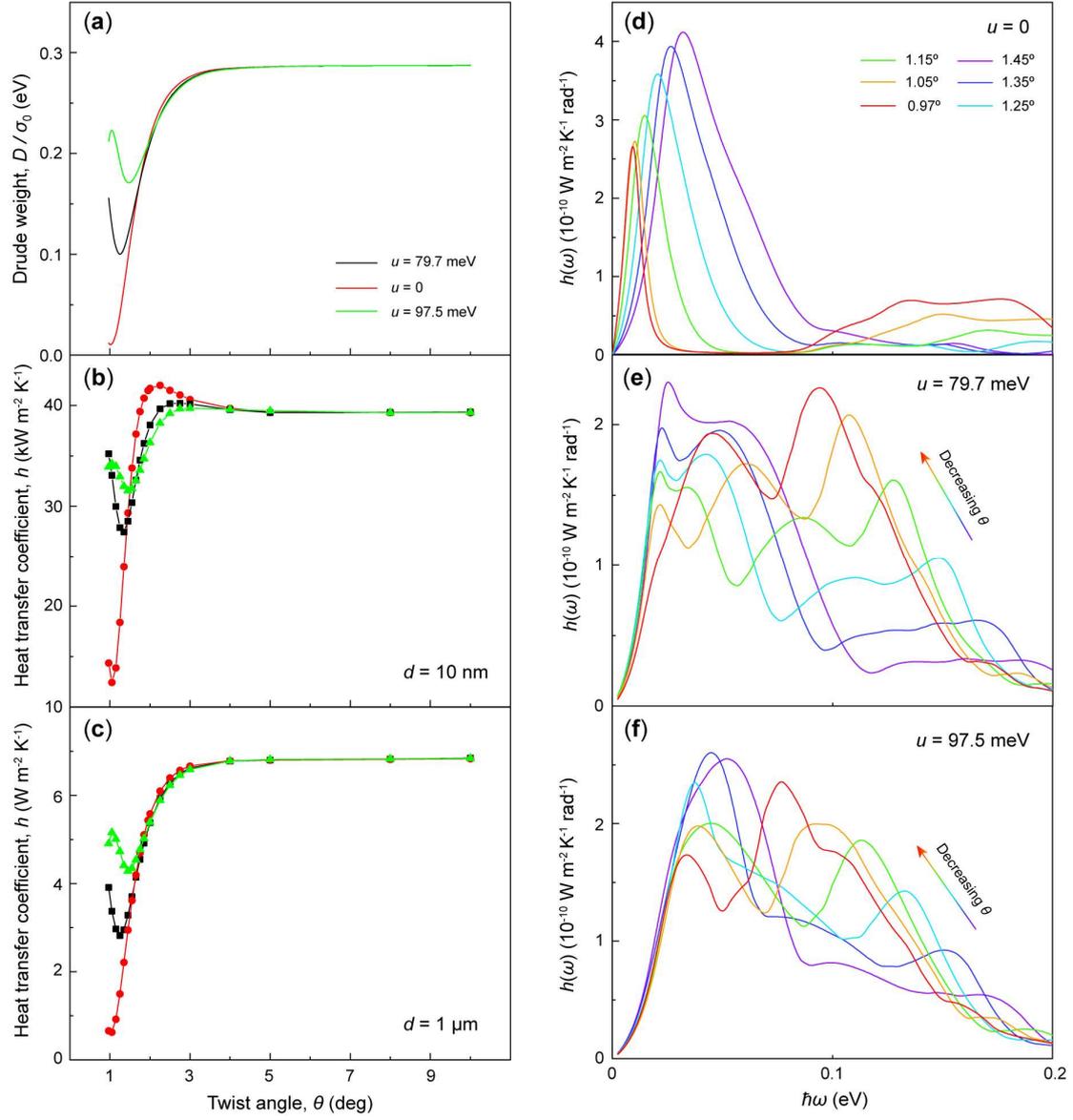

FIG. 7. The effect of lattice relaxation as modeled by varying the interlayer coupling energy $u$ at $T = 300$ K and $\mu = 0$. (a) Drude weight as a function of the twist angle. (b)-(c) The corresponding heat transfer coefficients for $d = 10$ nm and 1 μm, respectively. (d)-(f) The spectral heat transfer coefficient across a 10 nm gap for $u = 0$, 79.7 meV, and 97.5 meV, respectively.